\documentclass[microarticle]{elsarticle}
\usepackage[T1]{fontenc}
\usepackage{textcomp}
\usepackage{lineno}
\modulolinenumbers[5]
\usepackage{multirow}
\usepackage{soul}
\usepackage{xcolor}
\usepackage{amsmath,amsfonts,amssymb,amsthm}
\journal{arXiv}
\usepackage{enumerate}
\usepackage{textcomp}
\newtheorem{definition}{Definition}
\newtheorem{theorem}{Theorem}
\newtheorem{comment}{Comment}
\newtheorem{statement}{Statement}

\begin{document}
%%%%%%%%%%%%%%%%%%%%%%%%%%%%%%%%%%%%%%%%
\begin{frontmatter}
\title{Non-holonomic constraints: Considerations on the least action principle also from a thermodynamic viewpoint}
%\tnoteref{mytitlenote}}


%\author[mymainaddress,mysecondaryaddress]{Giulia Grisolia}

%\author[mysecondaryaddress]
\author[mymainaddress]{Umberto Lucia\corref{mycorrespondingauthor}}
\cortext[mycorrespondingauthor]{Corresponding author}
\ead{umberto.lucia@polito.it}

\author[mymainaddress]{Giulia Grisolia}

\address[mymainaddress]{Dipartimento Energia \textquotedblleft Galileo Ferraris\textquotedblright, Politecnico di Torino\\ Corso Duca degli Abruzzi 24, 10129, Torino, Italy}


\begin{abstract}
	The principle of least action seems not to lead to equations describing the motion consistent with the physical behaviour, for non-holonomic constraints. Here, a response is proposed for this fundamental problem in Mathematical Physics. Some considerations are also developed, based on the first and second law of thermodynamics.
\end{abstract}


\end{frontmatter}
%%%%%%%%%%%%%%%%%%%%%%%%%%%%%%%%%
%\linenumbers

\section{Introduction}
The Principle of Least Action represents  one of the most studied bases of Physics, and some controversies emerged during its historical development. 

In two papers, dated 1741 and 1744, submitted to the French Academy of Sciences, Pierre-Louis Moreau de Maupertuis (1698-1759) \cite{PLM1} mentioned a \textit{Principe de la moindre quantite d'action} (principle of least action), that he defined universal. In particular he highlighted that when a change occurs
in Nature, the Action required for that change is as small as possible. Then, Leonhard Euler (1707-1783) improved this result by pointing out that the true trajectory of any moving mass particle is the one, from among all possible trajectories connecting the same end point, that minimizes the action, which he realized to be the time integral of the twice kinetic energy \cite{eulero}. 

In the development of Mechanics, the method of least squares, introduced by Carl Friedrich Gauss (1777–1855), appeared first in the analysis of the elliptical orbit of the asteroid Ceres \cite{gauss}, strictly related to the Legendre's approach. The improvement of this approach led to the formulation that, from all possible motions, the actual one leads under given conditions to the least constraint, principle strictly related to the d’Alembert's Principle (Jean d’Alembert, 1717-1783). In this context, a statistical mechanical analysis of the Gauss's Principle has been developed in relation to its application for holonomic (constraints depend only on co-ordinates) and non-holonomic (non-integrable constraints on velocity) constraints, pointing out that the Gauss's principle is limited to arbitrary holonomic constraints and apparently, to nonholonomic constraint functions which
are homogeneous functions of the momenta \cite{DEGM}.

Nowadays, the principle of least action is used in Physics, even if it is often known as Hamilton's principle more than Maupertuis's principle, while it doesn't find interest in engineering, where some variational principles are used for applications \cite{Ramm}, even if a interest for the use of least action principle is growing in biomechanics and robotics \cite{biomech}. The variational methods are fundamental in the development of modern analytical mechanics \cite{flannery}. But, Flannery pointed out \cite{blochfirst,bloch,blochlast,flannery}:
\begin{quotation}
	``\textit{The least action principle can be applied only to holonomic and linear non-holonomic constraints, while it is not useful to obtain the correct equations of motion for general non-holonomic constraints}''.
	\item 
\end{quotation}
The problem pointed out by this statement can be summarised as the following question: Is it possible to prove that the principle of least action cannot be applied to general non-holonomic constraints?

In this paper, the problem of validity of the least action principle is discussed in order to propose a proof for answering to the question pointed out by Flannery. To do so, in Materials and Methods section the holonomic and non-holonomic constraints are discussed, in the Results section a proof in relation to the Flannery question is proposed, and in the last section some considerations are developed from a Thermodynamic viepoint.

%%%%%%%%%%%%%%%%%%%%%%%%%%%%%%%%%%%%%%%%%%%
\section{Materials and Methods}
%%%%%%%%%%
\subsection{Preliminary considerations}
The mechanical systems movements are restricted by constraints, due to material achievements which can be geometrically represented by lines, curves, planes, surfaces \cite{morro}. The constraint are described both by their mathematical equations and by the forces related to the binding reaction (physical description) \cite{morro}.

In order to determine the spatial position of a system of  $N$ material points,  the values of $N$ position vectors $\mathbf{r}$ are required, i.e. \cite{landau}:
\begin{itemize}
	\item $3N$ coordinates $r_i, i\in[1,3 ]$, if the system is free, i.e.,  the points representing the system are all possible;
	\item $n\leq 3N$ coordinates, if some of the points representing the system are subjected to restriction, and $n$ is the number of system's freedom degrees.
\end{itemize}
Consequently, for a general approach, it is convenient to introduce a system of $n \leq 3N$ (with $n = 3N$, if the system is free) generalized coordinates $q$ appropriate to the problem considered \cite{landau}. In order to determine the mechanical condition of a system at a given time, the values of the generalized velocities is also required: $(\mathbf {q},\dot{\mathbf{q}})$, where $\mathbf{q} = (q_1,\dots,q_n) $ and $\dot{\mathbf{q}} = (\dot{q} _1,\dots,\dot {q} _n)$, at the same time \cite{landau}.

In this context, the definition of constraint must be introduced, as follows:
\begin{definition} \emph{\cite{morro}} - 
	A constraint is said:
	\begin{enumerate}
		\item Holonomic, if any restriction on the possible configurations of the system follows the condition:
		\begin{equation}
			f(\mathbf{q},t)=0
			\label{olonomi}
		\end{equation}
		which is an integrable relation;
		\item Non-holonomic, if any restriction on the movements possible \emph{\cite{morro}} of the system follows the condition:
		\begin{equation}
			g(\mathbf{q};\dot{\mathbf{q}},t)=0
			\label{anolonomi}
		\end{equation}
		which is not an integrable relation.
	\end{enumerate}
	If the nonholonomic constraints represents a holonomic constraint, then it is integrable.
\end{definition}

%%%%%%%%%%%%%%%%%%%%%%%%
\subsection{Holonomic and Lagrangian systems}
The concept of map is fundamental to represent the domain of a real open set, thus its definition must be introduced.
\begin{definition} \emph{\cite{morro}}
	- 
	Let $Q$ be a set of points. It is said a map of size $n$ on $Q$ an application injective $\varphi: U \subseteq Q \rightarrow \mathbb{R}^n $, with image the open set $\varphi(U)$ in  $\mathbb{R}^n$. The $n$ functions $Q_ {i}:U \rightarrow \mathbb{R}, i\ in [1,n] $, such that $\forall x \in U: \varphi(x) = \big(Q_ {1}(x),\dots,Q_{n}(x)\big)$, are the coordinates associated with the fold $ \varphi$. The $\mathbf{q} = \{Q_ {i}\}_{i \in [1,n]}$ form a local coordinate system on all $Q$. It denotes the fold with the pair $(U,\varphi)$ or $(U,\mathbf{q}) $.
\end{definition}

In this context, the transition functions can be defined. They are applications between the two open systems $\mathbb{R}^n$, represented by functions as  $q_{1i}=\varphi_{12i}(q_{1h})$ and $q_{2i}=\varphi_{21i}(q_{2h})$, useful to describe a change in coordinates between one map to another one.
\begin{definition} 
	- \emph{\cite{morro}}
	Two maps of dimension $n$, $\varphi_{1}: U_{1}\rightarrow\mathbb{R}^{n}$ and $\varphi_{2}: U_{2}\rightarrow\mathbb{R}^{n}$, are said $C^{k}-$compatible if $U_{1}\cap U_{2}=\emptyset$ or if, when $U_{1}\cap U_{2}\neq\emptyset$, the two following conditions occur:
	\begin{enumerate}
		\item The sets $O_{1}=\varphi_{1}(U_{1}\cap U_{2})$ and $O_{2}=\varphi_{2}(U_{1}\cap U_{2})$, imagine of the intersection of the two domain on the two maps, are open;
		\item The transition function $\varphi_{12}: O_{1}\rightarrow O_{2}$ and $\varphi_{21}: O_{2}\rightarrow O_{1}$, defined as $\varphi_{12}=\varphi_{2}\circ \varphi_{1}^{-1}$ and $\varphi_{21}=\varphi_{1}\circ \varphi_{2}^{-1}$, with $\varphi_{1}$ and $\varphi_{2}$ of class $C^k$, are restricted to the intersection $U_{1}\cap U_{2}$.
	\end{enumerate}
\end{definition}

\begin{definition} \emph{\cite{morro}}
	- 
	On the set $Q$, a set of compatible maps is defined as $\mathcal{A}=\{\varphi_{\alpha}: U_{\alpha}\rightarrow \mathcal{R}^{n}; \alpha\in\mathcal{I}\}$, with $\mathcal{I}$ set of indices with domains $U_{\alpha}$, which are an overlap of $Q$. A set $Q$ with atlas is said differential variety of dimension $n$. 
\end{definition}
\begin{comment}
If the atlas has all the possible maps, then it is said full or filled or maximum. A differential variety is a set with maximum atlas.
\end{comment}	
\begin{comment} \emph{\cite{morro}}
	- 
	An atlas allows a topology, so a differential variety is also a topological space.
\end{comment}

\begin{definition} \emph{\cite{morro}}
	- 
	A set of points $\{P_{\nu}, \nu\in\mathcal{B}\}$, is said holonomic if its space of configurations $Q$ has the structure of differentiable variety. Then, $Q$  is said variety of configurations. The dimension $N$ of $Q$ is said number of freedom degree of the system. The coordinates $q_{i}$ related to every maps of $Q$ are said Lagrangian coordinates.
\end{definition}

\begin{comment} \emph{\cite{morro}}
	- 
	$\forall \nu\in\mathcal{B}, \exists \mathbf{r}_{\nu}: Q\rightarrow E_{3}$, i.e. there exists an application which assigns the position vector $\mathbf{r}_{\nu}$ of the point $P_{\nu}$ to ay configuration od teh system: known the coordinates $q_i$ on $Q$ the applications $\mathbf{r}_{\nu}$ are vectorial functions $\mathbf{r}_{\nu}(q_i)$. Consequently, the velocity is $$\dot{\mathbf{r}}_{\nu}=\sum_{i}\dfrac{\partial\mathbf{r}_{\nu}}{\partial q_i}\dot{q}_i$$.
\end{comment}

\begin{definition}
	- \emph{\cite{morro}}
	The motion act of a holonomic system is a set of vectors ${(\mathbf{r}_{\nu},\dot{\mathbf{r}}_{\nu}), \nu\in\mathcal{B}}$ such that:
	\begin{equation}
		\left\{\begin{array}{ll}
			\mathbf{r}_{\nu}=\mathbf{r}_{\nu}(q_i)
			\\
			\dot{\mathbf{r}}_{\nu}=\sum_{i}\dfrac{\partial\mathbf{r}_{\nu}}{\partial q_i}\dot{q}_i
		\end{array}
		\right.
	\end{equation}
\end{definition}
\begin{comment} \emph{\cite{morro}}
	- 
	If $Q$ is the configuration variety, then the set of the action acts is the tangent variety $TQ$; indeed, the $\dot{q}_i$ of the motion acts are the components of a vector tangent to $Q$ on the coordinates $q_i$.
\end{comment}

\begin{definition} \emph{\cite{morro}}
	- 
	A holonomic system is a system of points, whose possible configurations in all times are a differentiable variety $\bar{Q}$ of dimension $n+1$, said space-time of the configurations, such that:
	\begin{enumerate}
		\item There exists a differentiable function $t:\bar{Q}\rightarrow\mathbb{R}$, which assigns to any configuration its time;
		\item This application is such that  $\forall t\in\mathbb{R}$ the set $Q_t$ of all the possible configurations at the time $t$ is a sub-variaty of dimension $n$;
		\item There exists a differentiable variety $Q$ of dimension $n$ and a diffeomorfism $\varphi : \mathbb{R}\times Q\rightarrow\bar{Q}$, such that in any $i\in\mathbb{R}$ it generates a diffeomorfism  $\varphi_{t}:Q\rightarrow Q_{t}:q\mapsto\varphi (t,q)$ between the variety $Q$ and the variety $Q_t$.
	\end{enumerate}
The integer $n$ is the number of degree of freedom, and the variety $Q$ is the reference configuration variety.
\end{definition}

A holonomic system is made of constrained or free points. In dynamics, the action of the constrain is a force of reaction, called the constrain reaction, on which constitutive conditions must be imposed. 

The smooth constrain is represented by the orthogonality between the constrain reaction and the constrain itself. For a holonomic system, for a forces configuration and system $\mathbf{F}_{\nu}$, applied to every motion act related to an assigned configuration, it corresponds a power $W=\sum_{\nu}\mathbf{F}_{\nu}\cdot\dot{\mathbf{r}}_{\nu}$ of the forces; if the forces system is:
\begin{enumerate}
	\item An active force system $\mathbf{F}_{a\nu}$, then force laws related to positions and velocities are imposed, obtaining the consequent virtual power of the active forces $W_{a}^{(v)}$
	\item A virtual motion acts with the constrain reaction system  $\mathbf{F}_{r\nu}$, then a virtual power of the reactive forces $W_{r}$ is considered. 
\end{enumerate}
Thus, it follows the definition:
\begin{definition} \emph{\cite{morro}}
	- 	A holonomic system is perfect, or with perfect constrain, if the virtual power of the reactive forces is zero, for all virtual motion act.
\end{definition}
	The virtual power of the active forces is a linear form of the components $\delta q_{i}$, whose coefficients are defined as Lagrangian forces or Lagrangian components of the active forces:
\begin{equation} 
	W_{a}^{(v)}=\sum_{\nu}\mathbf{F}_{a\nu}\cdot\delta\mathbf{r}_{\nu}=\sum_{\nu}\mathbf{F}_{a\nu}\cdot\sum_{i}\frac{\partial\mathbf{r}_{\nu}}{\partial q_{i}}\delta q_{i}=\sum_{i}\varphi_{i}\delta q_{i}
\end{equation}
from which
\begin{equation}
	\varphi_{i}=\sum_{\nu}\mathbf{F}_{a\nu}\cdot\frac{\partial \mathbf{r}_{\nu}}{\partial q_{i}}
\end{equation}
If the active forces are functions of the positions and of the velocities then the Lagrangian forces are $\varphi =\varphi (\mathbf{q},\dot{\mathbf{q}},t)$.

\begin{definition}
	- \emph{\cite{morro}}
	The dynamic state of a mechanical system is the time distribution of the positions, velocities and accelerations of the points of the system.
\end{definition}

A virtual power $W_{m}^{(v)}$ of the mass forces, named inertial forces too, is associated to each dynamic state.  The inertial forces are defined by the Newton Law $$\mathbf{F}_{m\nu}=-m_{\nu}\mathbf{a}_{\nu}$$
The virtual power of the inertial forces is a linear form of the components $\delta q_{i}$ of the virtual motion act, too:
\begin{equation}
	W_{m}^{(v)}=\sum_{\nu}\mathbf{F}_{m\nu}\cdot\delta\mathbf{r}_{\nu}=-\sum_{\nu}m_{\nu}\mathbf{a}_{\nu}\cdot\sum_{i}\frac{\partial\mathbf{r}_{\nu}}{\partial q_{i}}\delta q_{i}=\sum_{i}\tau _{i}\delta q_{i}
\end{equation}
from which it follows:
\begin{equation}
	\tau _{i}=-\sum_{\nu}m_{\nu}\mathbf{a}_{\nu}\cdot\frac{\partial\mathbf{r}_{\nu}}{\partial q_{i}}
\end{equation}

\begin{statement} \emph{\cite{morro}} - \emph{\textbf{Lagrange-D'Alembert Principle}} - 
	In any dynamic state of a system with perfect constrains, for all virtual motion acts, the sum of the virtual powers of the active forces and of the inertial forces equals zero:
	\begin{equation}
		W_{a}^{(v)}+W_{m}^{(v)}=0
	\end{equation}
\end{statement}

A system $(Q,\mathcal{L})$ is said Lagrangian, if it is a differential variety $Q$ of dimension $n$, said configuration variety, with an associated real function $\mathcal{L}:TQ\times\mathbb{R}\rightarrow\mathbb{R}$. If the system is time independent the Lagrangian is a function $$\mathcal{L}:TQ\rightarrow\mathbb{R}$$ 
	
	The Lagrangian dynamics is the set of curves expressed by the first order system of $2n$ differentiable equations \cite{morro}:
	\begin{equation}
		\left\{
		\begin{array}{ll}
			\dot{q}_{i}=\dfrac{dq_{i}}{dt}
			\\
			\dfrac{d}{dt}\bigg(\dfrac{\partial\mathcal{L}}{\partial\dot{q}_{i}}\bigg)-\dfrac{\partial\mathcal{L}}{\partial q_{i}}
		\end{array}
		\right.
		\label{eulero-lagrange}
	\end{equation}
	where the \emph{(\ref{eulero-lagrange})$_{2}$} equations are the Euler-Lagrange ones.

	For the holonomic systems the intrinsic properties of the Euler-Lagrange equations, i.e. the independence of the Lagrangian coordinates choose, is the consequence of application of the Lagrange-D'Alembert principle to a Lagrangian equation system.

	A functional is an application $\phi :\Omega\rightarrow\mathbb{R}$ such that for all $n-$tupla of functions corresponds a real number. A functional is differentiable in a point $q_{i}(t)\in\Omega$ if for all the chooses of the growth, said variations, $\delta q_{i}(t)\in\Omega$ there exists the following relation \cite{morro}:
	\begin{equation}
		\phi (q_{i}+\delta q_{i})=\phi (q_{i})+\delta\phi (q_{i},\delta q_{i})+\mathcal{R}
	\end{equation}
	where $\delta\phi$ is a linear functional of $\delta q_{i}$ and $\mathcal{R}$ is a functional of upper order in the same increases.

\begin{definition} \emph{\cite{morro}}
	- 
	A variation $\delta q_{i}(t)$ is said end fixed if:
	\begin{equation}
		\delta q_{i}(t_{1})=\delta q_{i}(t_{2})=0
		\label{estremifissi}
	\end{equation}
\end{definition}

\begin{definition} \emph{\cite{morro}}
	- 
	The action is defined as:
	\begin{equation}
		\mathcal{A}=\int_{t_{1}}^{t_{2}}\mathcal{L}\big(t,q_{i}(t),\dot{q}_{i}\big)dt
	\end{equation}
\end{definition}

\begin{theorem} \emph{\cite{landau}}
	- \emph{\textbf{Least action principle}}.\\
	The function $q_{i}(t)$, for which $\delta\mathcal{A}=0$ for all \textbf{end fixed variations}, are only the solutions of the differential system \emph{(\ref{eulero-lagrange})}, where the Lagrangian is defined up to a function of the coordinates and time.
\end{theorem}

A general approach to mechanical systems can be developed by using the \textit{least action principle}, named also \textit{Hamilton principle}, for which the mechanical system is described using a Lagrangian function $\mathcal{L}(\mathbf{q};\dot{\mathbf{q}},t)$ from which the action can be obtained \cite{landau}:
\begin{equation}
	\mathcal{A}=\int_{t_1}^{t_2}\mathcal{L}(\mathbf{q};\dot{\mathbf{q}},t)
\end{equation}
The Hamilton principle states that the motion of a system follows the path $\mathbf{q}(t)$ for which the action is minimum:
\begin{equation}
	\delta\mathcal{A}=\delta\int_{t_1}^{t_2}\mathcal{L}(\mathbf{q};\dot{\mathbf{q}},t)=0
\end{equation}

The proof of this relation can be obtained starting from the hypothesis that the least value of the action is $\mathbf{q}(t)$, and a small variation $\delta\mathbf{q}$ around it, are considered. Then for
\begin{equation}
	\mathbf{q}(t) + \delta\mathbf{q}(t)
	\label{posizione}
\end{equation}
the action $S$ increases \cite{landau}, but for $t=t_1$ and $t=t_2$ the relation (\ref{posizione}) must have the fixed values $\mathbf{q}(t_1)=\mathbf{q}_1$ and $\mathbf{q}(t_2)=\mathbf{q}_2$; this statement represent \textit{fundamental condition} for the Hamilton principle \cite{landau}:
\begin{equation}
	\delta\mathbf{q}(t_1)=\delta\mathbf{q}(t_2)=0
	\label{condizione1}
\end{equation}

Consequence of the least action principle is the Lagrange differential equations:
\begin{equation}
	\frac{d}{dt}\frac{\partial\mathcal{L}}{\partial \dot{q}_i}-\frac{\partial\mathcal{L}}{\partial q_i}=0 \qquad i\in[1,n]
\end{equation}

%%%%%%%%%%%%%%%%%%%%
\subsection{Non-holonomic constrains}
A free point $P$, from any initial position $P_{0}$, at the initial time $t_{0}$, can move of an elementary displacement $dP=\mathbf{v}dt$; for a constrained point, these displacements are confined due to the constrain \cite{civita1}. A holonomic system in a initial configuration at the time $t_0$, can have a transition to another configuration at the time $t_0+dt$, infinitely near to the initial one \cite{civita1}.

\begin{definition} \emph{\cite{civita1} -  }
	A possible displacement at the time $t$, starting from a configuration $C$, is any infinitesimal displacement of a honolomic system, which allows it to have a transition from the configuration $C$ at the time $t$ to a new  configuration $C'$ at the time $t+dt$:
	\[
	P_i=P_i(\mathbf{q};t)\mapsto P_i+dP_i=P_i(\mathbf{q}+d\mathbf{q};t+dt)
	\]
	from which the possible displacement are the $n$ equations:
	\begin{equation}
		dP_i=\sum_{k}\frac{\partial P_i}{\partial q_k}dq_k+\frac{\partial P_i}{\partial t}dt=\nabla_{\mathbf{q}} P_{i}\cdot d\mathbf{q}+\frac{\partial P_i}{\partial t}dt
	\end{equation}
\end{definition}

If the virtual displacement are coupled the holonomic constrains equations (\ref{olonomi}), related to the displacements themselves, represented by the $l$ equations, it follows:
\begin{equation}
	df_j=\sum_{k}\frac{\partial f_j}{\partial q_k}dq_k+\frac{\partial f_j}{\partial t}dt=\nabla_{\mathbf{q}}  f_{j}\cdot d\mathbf{q}+\frac{\partial f_j}{\partial t}dt=0
\end{equation}
and only $n-l$ free Lagrangian coordinates can be obtained. Dividing for $dt$ the non-holonomic constrain relation can be obtained (\ref{anolonomi}):
\begin{equation}
	\frac{df_j}{dt}=\sum_{k}\frac{\partial f_j}{\partial q_k}\dot{q}_k+\frac{\partial f_j}{\partial t}=\nabla_{\mathbf{q}}  f_{j}\cdot\dot{\mathbf{q}}+\frac{\partial f_j}{\partial t}=\sum_{k}a_{jk}\dot{q}_k+b_j=(\mathbf{a}\cdot\dot{\mathbf{q}}+b)_j=0
	\label{anol}
\end{equation}
It is a restrain in the motion. 

So, the displacement are limited and the virtual displacements must be introduced:
\begin{definition} \emph{\cite{civita1} - }
	A virtual displacement is any hypothetical displacement, which allows the system to have a transition from a configuration $C$ to another infinitesimal near one $C'$, allowed by the constrains at the same time.
\end{definition}

Consequently, for the non-holonomic constrains $dt=0$, and the relation (\ref{anol}) becomes:
\begin{equation}
	\nabla_{\mathbf{q}}  f_{j}\cdot d\mathbf{q}+\frac{\partial f_j}{\partial t}dt=0\Rightarrow\mathbf{a}\cdot\delta{\mathbf{q}}=0
	\label{anol1}
\end{equation}

	The live force is the value of the kinetic energy, and the following theorem can be introduced:
\begin{theorem} \emph{\cite{civita2} - }
	\emph{\textbf{Theorem of live forces or of K\"onig}} - The live force of any system in motion is the sum of the live force of the centre of mass, and the one of the motion in relation to the centre of mass.
\end{theorem}

Volterra pointed out that the Lagrangian is an explicit function of $\dot{\mathbf{q}}$ \cite{civita2}.

%%%%%%%%%%%%%%%%%%%%%%%%%%%%%%%%
\section{Results}
%%%%%%%%%%%%%%%%%%%%%%%%
Non-holonomic systems, term coined by Heinrich Rudolf Hertz (1857-1894) in 1894, are mechanical systems with constraints on their velocity
that are not derivable from position constraints.

There are some differences between non-holonomic and Hamiltonian or Lagrangian systems, e.g. \cite{bloch}:
\begin{itemize}
	\item Non-holonomic systems arise from the Lagrange-d'Alembert principle and not from Hamilton's principle;
	\item Energy is always preserved, while momentum is not always preserved;
	\item Their volume in the phase space may not be preserved.
\end{itemize}

The equations of motion of a non-holonomic system in the form of the Euler-Lagrange equations, with the correction obtained by introducing some additional terms related to the constraints, but without Lagrange multipliers, when some of the configuration variables are cyclic, was obtained in 1895 by Sergej Alekseevi\u{c} \u{C}aplygin (1869-1942), who realised also the importance of an invariant measure in non-holonomic dynamics \cite{bloch}.

A fundamental question on the non-validity of the principle of least action for non-holonomic constraints, highlighted by Flannery \cite{flannery}, is suggested, based on the previous definitions and theorems. In this paper, a response \cite{lucia1,lucia2} to this question is proposed.

The basis of the least action principle is the evaluation of the variations under the hypothesis of the fix ends (\ref{estremifissi}) \cite{elsgolts}. 

For non-holonomic constrains (\ref{anol}) and (\ref{anol1}), at least, one of the virtual displacements can be written as a linear combination of the others; i.e., 
\begin{equation}
	\delta q_i(t)=a_{i}^{-1}\sum_{j}a_{ij}\delta q_j
\end{equation}
so, the relations (\ref{anol}) e (\ref{anol1}), fundamental for the validity of the least action principle, are not satisfied. 

Consequently, for non-holonomic constrains, the fundamental conditions for the use of the least action principle are not verified, proving that for a general non-holonomic constraint the principle of least action cannot be used. An alternative approach from thermodynamics is also suggested.

%%%%%%%%%%%%%%%%%%%%%%%%%%%%%%%%%
\section{Discussion and Conclusions}
%%%%%%%%%%%%%%%%%%%%%%%%%
The proof proposed limits the use of the principle of least action to the holonomic and linear non-holonomic constraints \cite{flannery}. But, it is important to find an alternative approach for generic non-holonomic systems. To do so, some considerations can be introduced from thermodynamics, with particular regards to the second law.

A thermodynamic system is a physical system, which interacts with its environment, by exchanging heat and work \cite{borel}. For such system, it is possible to write the kinetic energy theorem in the following form \cite{houberects,cali,giaretto}:
\begin{equation}
	\delta W_{es} + \delta W_{fe} + \delta W_{i} = dE_k
\end{equation}
where $\delta$ represents the pathway dependent differential, $W_{es}$ is the work done by external forces on the border of the system, $W_{fe}$ is the work lost due to external irreversibility, $E_k$  is the kinetic energy of the system, and $W_{i}$ is the internal work, such that:  
\begin{equation}
	\delta W_i = \delta W_{i}^{rev} - \delta W_{fi}
\end{equation}
where $rev$ indicates the reversible (ideal) internal work and $W_{fi}$ depicts the work lost due to internal irreversibility. Considering the relation \cite{houberects,cali,giaretto}:
\begin{equation}
	\delta W_{se} + \delta W_{es} + \delta W_{fe} = 0
\end{equation}
where $W_{se}$ is the the work by the internal forces on the border of the system towards the outside of the system. As a consequence of this approach, the first law of thermodynamics appears in the following form:
\begin{equation}
	\delta Q - \delta W_{se} = dU + dE_k
\end{equation}

Now, defining the Lagrangian as \cite{landau,morro}:
\begin{equation}
	\mathcal{L} = E_k - E_p
\end{equation}
where $E_p = W_i + W_{es}$ is the potential energy. Consequently, it follows:
\begin{equation}
	\mathcal{L}  = E_k - ( W_i +  W_{es}) =  W_{fe}
\end{equation}

Considering the Gouy-Stodola theorem \cite{houberects,cali,giaretto}:
\begin{equation}
	W_{fe} = -T_0 \int_0^\tau \Sigma \cdot dt
\end{equation}
where $T_0$ is the environmental temperature, $\Sigma$ is the entropy generation rate, and $t$ is the time. Considering the duration of a process $\tau$, i.e., the time in which a process occurs, and the mean value of the entropy generation rate $\bar{\Sigma}$, it is possible to obtain the entropy generation, $S_g$, for any real process by using an engineering thermodynamic approach, as follows:
\begin{equation}
	S_g = \bar{\Sigma}\cdot \tau
\end{equation}
Consequently, the Lagrangian results:
\begin{equation}
	\mathcal{L} = - T_0\, S_g
\end{equation}
and the action $\mathcal{A}$ results:
\begin{equation}
	\mathcal{A} = -T_0 \int S_g dt
\end{equation}
This last relation is very interesting because the entropy generation results always integrable for a real process, as usually done in engineering thermodynamics, independently on the formal expression of the displacements, obtaining:
\begin{equation}
	\mathcal{A} = - T_0\, \bar{S}_g \, \tau
\end{equation}
consequently, the least action can be used by evaluating the maximum entropy generation, which is related only to dissipation. In this way, the analysis of the motion for non-holonomic systems becomes the analysis of the dissipation during the motion.

These considerations represent a starting point in the analysis of non-holonomic constraints, proposing a thermodynamic viewpoint, which analytically confirms, on the bases of the first and second law of thermodynamics, the considerations highlighted in Ref. \cite{entr}.

%%%%%%%%%%%%%%%%%%%%%%%%%%%%%%%%%%
%\section*{References}
\bibliography{References}

\end{document}